\documentclass[a4paper,keeplastbox, nospread]{jacow}
\makeatletter%
	\ifboolexpr{bool{xetex}}
	 {\renewcommand{\Gin@extensions}{.pdf,%
	                    .png,.jpg,.bmp,.pict,.tif,.psd,.mac,.sga,.tga,.gif,%
	                    .eps,.ps,%
	                    }}{}
\makeatother

\ifboolexpr{bool{xetex} or bool{luatex}} 
 {}                                      
 {\usepackage[utf8]{inputenc}}           

\usepackage[final]{pdfpages}
\usepackage{multirow}

\ifboolexpr{bool{jacowbiblatex}}%
 {%
  \addbibresource{jacow-test.bib}
  \addbibresource{biblatex-examples.bib}
 }{}
\newcommand{\lum}{cm$^{-2}$s$^{-1}$~}
\newcommand{\slum}{cm$^{-2}$s$^{-1}$/mA$^2$}

\begin{document}

\title{Highlights from \NoCaseChange{SuperKEKB} Phase 2 Commissioning} 

\author{Y. Ohnishi\thanks{yukiyoshi.onishi@kek.jp}, KEK, 1-1 OHO, Tsukuba, Ibaraki 305-0801, Japan \\
on behalf of the SuperKEKB Commissioning Group and the Belle II Commissioning Group} 
	
\maketitle

\begin{abstract}
SuperKEKB is an electron-positron asymmetric-energy collider to search new physics phenomena appeared in B-meson decays. 
In order to accomplish this purpose, 40 times the luminosity as high as the KEKB collider is demanded. 
The strategy is that the vertical beta function at the IP is squeezed down to 1/20 and the beam currents double those of KEKB 
while keeping the same beam-beam parameter. 
The vertical beta function at the interaction point(IP) will be much smaller than the bunch length, 
however, the hourglass effect which degrades the luminosity will be reduced by adopting a novel ``nano-beam'' scheme. 
First of all, the Phase 2 commissioning was focused on the verification of nano-beam scheme. 
Secondary, beam related background at the Belle II detector was also studied for the preparation of the pixel vertex detector 
installed before the Phase 3 operation. 
The preliminary results and accomplishments of the commissioning in Phase 2 will be reported in this article.
\end{abstract}
\section{Introduction}
SuperKEKB is an electron-positron collider\cite{ref:SuperKEKB} and the Belle II detector\cite{ref:BelleII}
built to explore new phenomena in particle physics.
The physics program of the next B-factory delivering ultra high statistics is almost independent of, and complementary to,
the  high energy experiments at the LHC.
The target luminosity is 8$\times$10$^{35}$ cm$^{-2}$s$^{-1}$, which is 40 times the performance of the predecessor, 
KEKB\cite{ref:KEKB}, which has been operated for 11 years until 2010.
The strategy for the luminosity upgrade is a nano-beam scheme.
The nano-beam scheme was first proposed by P. Raimondi in Italy\cite{ref:SuperB}. 
The collision of low emittance beams under a large crossing angle allows squeezing the beta functions at the IP
much smaller than the bunch length.
Consequently, extremely higher luminosity can be expected with only twice the beam current of KEKB.

The SuperKEKB operation is divided by 3 stages, Phase 1, Phase 2, and Phase 3.
The upgrade work was started after the shutdown of KEKB, and it took 6 years to make the Phase 1 commissioning ready.
The final focus system(QCS)\cite{ref:QCS} and Belle II detector were not installed in Phase 1\cite{ref:Phase1}. 
The subjects were vacuum scrubbing for new vacuum system replaced with ante-chambers, 
low emittance tuning for new arc lattice to realize low emittance, and beam background study 
prepare for the installation of Belle II detector before Phase 2.
The final focus system and Belle II detector were installed during a long shutdown between Phase 1 and Phase 2.
Prior to the main ring operation, the commissioning of the positron damping ring\cite{ref:DR} started on 8th February 2018 
almost in 2 years after the Phase 1 commissioning.
The Phase 2 commissioning started on 19th March 2018.
The commissioning in Phase 2 was finished on 17th July 2018 and the duration was about 4 months in total.
The common machine parameters during Phase 2 are shown in Table~\ref{tab:machine-1}.

The Phase 3 operation will start in the early 2019, which is a full-scale collider experiment 
after installation of the pixel vertex detector(PXD) to Belle II.
\begin{table}[hbt]
   \centering
   \caption{
     Machine Parameters related to the RF system in Phase 2.
     The intra-beam scattering and other collective effects are not included.
   }
   \begin{tabular}{lccc}
       \toprule
       & \textbf{LER} & \textbf{HER} & Unit \\
       \midrule
       Beam Energy    & 4                    &  7                 & GeV \\
       Circumference  & \multicolumn{2}{c}{3016.3} & m  \\ 
       Harmonic no.   & \multicolumn{2}{c}{5120} & \\ 
       Total RF voltage     & 8.4                    &  12.8                 & MV \\
       $\alpha_p$     & 2.88$\times$ 10$^{-4}$ & 4.50$\times$ 10$^{-4}$ & \\
       $\sigma_z$     & 4.8                  & 5.4                   & mm \\
       $\sigma_\delta$ & 7.53 $\times$ 10$^{-4}$ & 6.3 $\times$ 10$^{-4}$ & \\
       $U_0$          & 1.76                 & 2.43                  & MV \\
       $\nu_s$        & -0.0220               & -0.0258                & \\
       \bottomrule
   \end{tabular}
   \label{tab:machine-1}
\end{table}
\section{Target of the Phase 2 Commissioning}
The overlap region for the narrow colliding beams with a large crossing angle can be small along the beam axis which implies
a head-on collision of effective beams having the very short bunch length.
A picture of the effective beams is a projection of the real beams to the $x$-axis which is an isovolumetric deformation.
The effective beam is considered in the nano-beam scheme which is written by
\begin{eqnarray}
\sigma_{z,eff} &=& \frac{\sigma_x^*}{\phi_x} \\
\sigma_{x,eff}^* &=& \sigma_z\phi_x,
\end{eqnarray}
where $\sigma_x^*$ is the horizontal beam size at the IP, $\sigma_z$ is the bunch length, and $\phi_x$ is the half crossing angle.
Then, the luminosity and beam-beam parameters are calculated by using the effective beam.
In order to avoid an hourglass effect, the following condition is necessary.
\begin{eqnarray}
\beta_y^* \ge \sigma_{z,eff} = \frac{\sigma_z}{\Phi},
\end{eqnarray}
where the Piwinski angle is defined by
\begin{eqnarray}
\Phi=\frac{\sigma_{x,eff}^*}{\sigma_x^*}.
\end{eqnarray}
The Piwinski angle is larger than 10 in the nano-beam scheme while that of the conventional collision scheme is
smaller than 1.
Therefore, $\beta_y^*$ can be squeezed down to 300 $\mu$m with assuming $\sigma_z$ = 6 mm and $\Phi$ = 20.
The arc cell and the interaction region are designed to realize the low emittance and large Piwinski angle in the nano-beam scheme.
Another point of view of the overlap region is distributions of primary vertex positions.
In the case of nano-beam scheme, the vertex distribution along the $z$-axis is constrained in the small region, 
for example $\sigma_{vertex}$ = 550 $\mu$m in Phase 2 
in contrast with $\sigma_{vertex}$ = 4.5 mm in the conventional scheme such as KEKB which are measured by the vertex detectors.

The luminosity in the nano-beam scheme is written by
\begin{eqnarray}
L=\frac{N_-N_+n_bf_0}{4\pi\sigma_{x,eff}^*\sqrt{\varepsilon_y\beta_y^*}}\simeq\frac{\gamma_\pm}{2er_e}\frac{I_\pm\xi_{y\pm}}{\beta_y^*},
\label{eq:lum}
\end{eqnarray}
where the vertical beam-beam parameter is
\begin{eqnarray}
\xi_{y\pm}=\frac{r_eN_\mp}{2\pi\gamma_\pm\sigma_{x,eff}^*}\sqrt{\frac{\beta_y^*}{\varepsilon_y}}.
\label{eq:beambeam}
\end{eqnarray}
In Eqs.~\ref{eq:lum} and \ref{eq:beambeam}, $\varepsilon_{y-}=\varepsilon_{y+}$ and $\beta_{y-}^*=\beta_{y+}^*$ are assumed,
$n_b$ is the number of bunches, $N_\pm$ is the number of particles in a bunch, $f_0$ is the revolution frequency.
When the $\beta_y^*$ is squeezed, the $\xi_y$ becomes small proportional to $\sqrt{\beta_y^*}$,
however, the luminosity increases proportional to $1/\sqrt{\beta_y^*}$.
If we can make $\varepsilon_y$ small similar to the ratio of the beta squeezing, the luminosity is proportional to $1/\beta_y^*$
with keeping the same $\xi_y$.

The targets in the Phase 2 commissioning are
\begin{enumerate}
\item Verification of the nano-beam scheme. Confirm the luminosity increases even though $\beta_y^*$ becomes smaller than $\sigma_z$.
      The beam-beam parameter is $\xi_y$ $>$ 0.03.
      The luminosity is $L$ = 10$^{34}$ cm$^{-2}$s$^{-1}$ at 1 [A] in LER.
\item Understanding and reduction of Belle II beam related backgrounds.
\item Establishment of the injection system\cite{ref:Injector}.
\end{enumerate}
\section{Achievements in the Phase 2 Commissioning}
The commissioning of the HER and LER was started with the large beta functions at the IP, 
which is called ``detuned optics'', in order to capture beams.
The beta functions at the IP were $\beta_x^*$ = 400 mm and $\beta_y^*$ = 81 mm in the HER,
$\beta_x^*$ = 384 mm and $\beta_y^*$ = 48.6 mm in the LER.
After beams were stored first, calibrations of the QCS response, beam-based alignments, and vacuum scrubbing were performed.

\begin{figure}[htb]
   \centering
   \includegraphics[width=240pt]{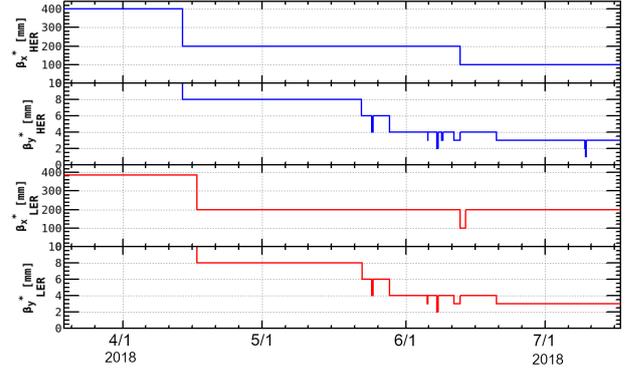}
   \caption{History of the beta squeezing at the IP.}
   \label{fig:BetaSqueeze}
\end{figure}

The beta functions at the IP were squeezed down to 200 mm for $\beta_x^*$ and 8 mm for $\beta_y^*$ for each ring
in the middle of April 2018.
Then, the beta functions at the IP were squeezed down to 3 mm for $\beta_y^*$ gradually.
The history of beta squeezing is shown in Fig.~\ref{fig:BetaSqueeze}.
The smallest $\beta_y*$ is 1.5 mm in the HER and 2 mm in the LER, which were tests to squeeze $\beta_y^*$ 
and the global optics correction was applied although those were not used for the luminosity run.
This value is the smallest $\beta_y^*$ in the world.

\begin{figure*}[htb]
	\centering
	\includegraphics[width=400pt]{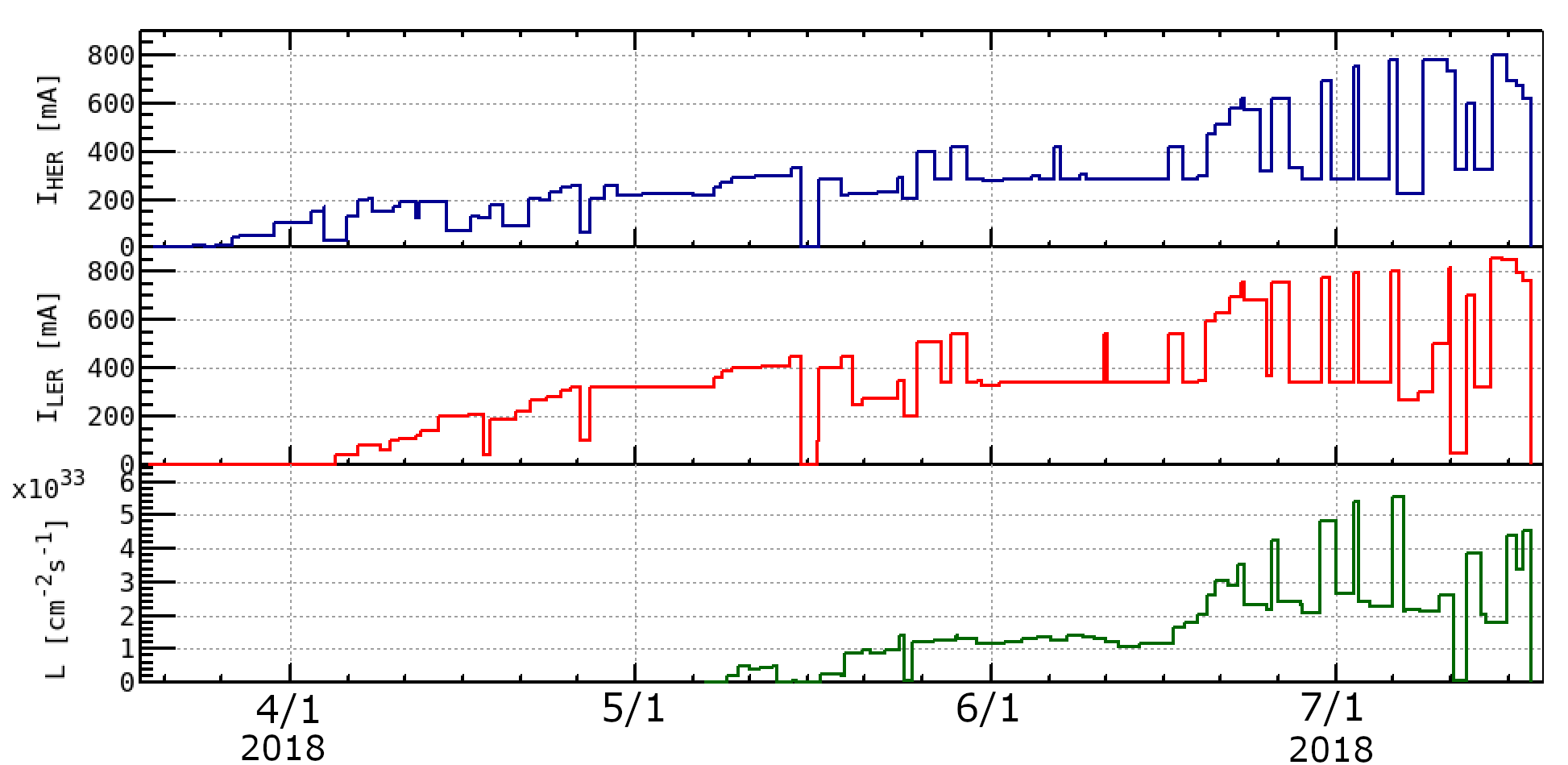}
	\caption{
	  History of beam currents and luminosity in the Phase 2 commissioning. }
	\label{fig:history}
\end{figure*}

The maximum beam current is 860 mA in the LER and 800 mA in the HER during Phase 2, respectively. 
The history of beam currents and luminosity in the Phase 2 commissioning is shown in Fig.~\ref{fig:history}. 

\subsection{Luminosity Performance}
\begin{figure}[htb]
   \centering
   \includegraphics[width=240pt]{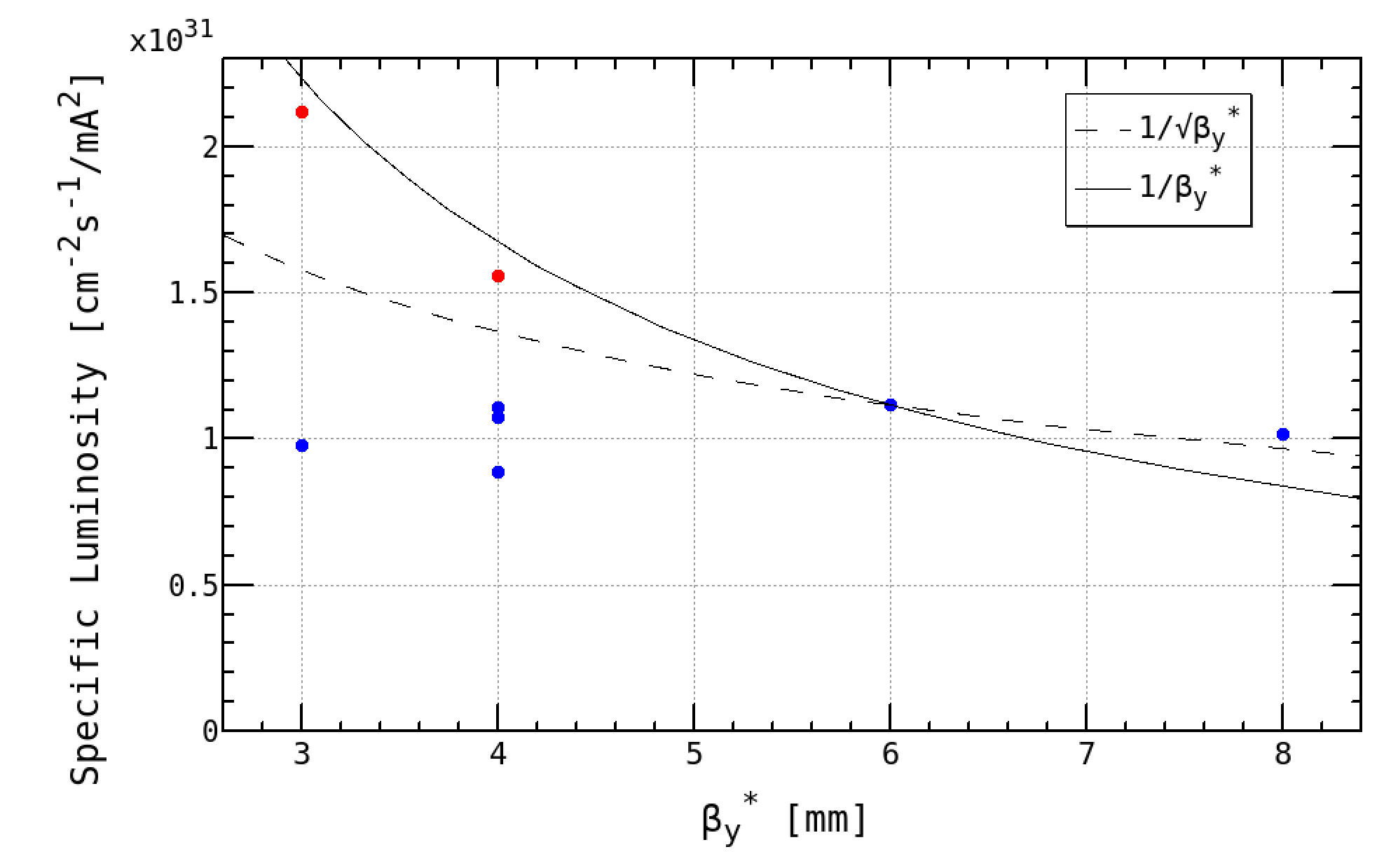}
   \caption{Specific luminosity as a function of $\beta_y^*$. 
     Red plots indicate the specific luminosity after the correction of XY couplings at the IP and the waist with QC1s.}
   \label{fig:specLum-by}
\end{figure}

Figure~\ref{fig:specLum-by} shows the specific luminosity as a function of $\beta_y^*$. 
The specific luminosity is defined by
\begin{eqnarray}
L_{sp} &=& \frac{L}{n_bI_{b+}I_{b-}}=\frac{1}{4\pi\sigma_z\phi_xe^2f_0\bar{\sigma}_y^*} \nonumber \\ 
      &=& \frac{1.25\times10^{25}}{\bar{\sigma}_y^*}~~~[cm^{-2}s^{-1}/mA^2],
\label{eq:speclum}
\end{eqnarray}
where $I_{b\pm}$ is the bunch current and
\begin{eqnarray}
\bar{\sigma}_y^*=\frac{\sqrt{\sigma_{y-}^{*2}+\sigma_{y+}^{*2}}}{\sqrt{2}}
=\frac{\Sigma_y^*}{\sqrt{2}}.
\end{eqnarray}
When $\beta_y^*$ was squeezed from 6 mm to both 4 mm and 3 mm which were smaller than the bunch length,
the specific luminosity was not improved in the early luminosity tuning. 
The global optics correction has been successfully working\cite{ref:opticsCorrect}.
The vertical dispersions and XY couplings were corrected by using skew quadrupole coils wound in the sextupole magnets.
The beta functions and horizontal dispersions were corrected by adjustment of the quadrupole field gradient and
horizontal local bump orbit at the sextupole pairs.
The typical result of global optics correction is shown in Table~\ref{tab:optics-correct}.
We suspected that there is machine error locally in the vicinity of the IP 
such as a waist shift, local XY coupling at the IP, and so on\cite{ref:R2atIP}.
The machine error due to the QCS can affect the vertical beam size at the IP as following:
\begin{eqnarray}
\sigma_y^{*2} &=&  \mu^2\varepsilon_y\left(\beta_y^*+\frac{\Delta s^2}{\beta_y^*}\right)+(\eta_y^*\sigma_\delta)^2  \nonumber \\
           & & + \varepsilon_x\frac{(r_2+r_4\Delta s)^2}{\beta_x^*}+\varepsilon_x\beta_x^*(r_1+r_3\Delta s)^2,
\end{eqnarray}
where $r_{1-4}$ are the XY coupling parameters, $\mu^2=1-(r_1r_4-r_2r_3)$, $\Delta s$ is the waist shift.
The physical coordinate system of a particle, $(x,p_x,y,p_y)$, is written by 
\begin{eqnarray}
\left(\begin{array}{c}
x \\ p_x \\ y \\ p_y
\end{array}\right)=
\left(\begin{array}{cccc}
\mu & 0 & r_4 & -r_2 \\
0 & \mu & -r_3 & r_1 \\
-r_1 & -r_2 & \mu & 0 \\
-r_3 & -r_4 & 0 & \mu
\end{array}\right)
\left(\begin{array}{c}
u \\ p_u \\ v \\ p_v
\end{array}\right),
\end{eqnarray}
where $(u,p_u,v,p_v)$ is the decoupled coordinate system.

The specific luminosity was improved by correction of $r_2$ which is one of the XY couplings with QCS skew quadrupole correctors
and the waist position with adjustment of the QCS quadrupole coils.
Thus, it is found that the specific luminosity is consistent with the behavior of $1/\beta_y^*$ 
even though $\beta_y^*$ is squeezed smaller than the bunch length.
\begin{table}[hbt]
   \centering
   \caption{
     The typical result of global optics correction.
     The value is rms for those at all BPMs in the ring.
     The $\varepsilon_y$ is the projected vertical emittance measured by the X-ray beam size monitor.
   }
   \begin{tabular}{lccc}
       \toprule
       \textbf{Item} & \textbf{LER} & \textbf{HER} & Unit \\
       \midrule
       rms($\Delta\beta_x/\beta_x$) & 2 & 3 & \% \\
       rms($\Delta\beta_y/\beta_y$) & 4 & 3 & \% \\
       rms($\Delta y/\Delta x$)     & 0.014 & 0.008 & \\
       rms($\Delta\eta_x$)          & 10 & 9 & mm \\
       rms($\Delta\eta_y$)          & 4 & 3 & mm \\
       $\varepsilon_y$              & 23 & 9 & pm \\
       $\varepsilon_y/\varepsilon_x$  & 1.35 & 0.20 & \% \\
       \bottomrule
   \end{tabular}
   \label{tab:optics-correct}
\end{table}

\begin{figure}[htb]
   \centering
   \includegraphics[width=240pt]{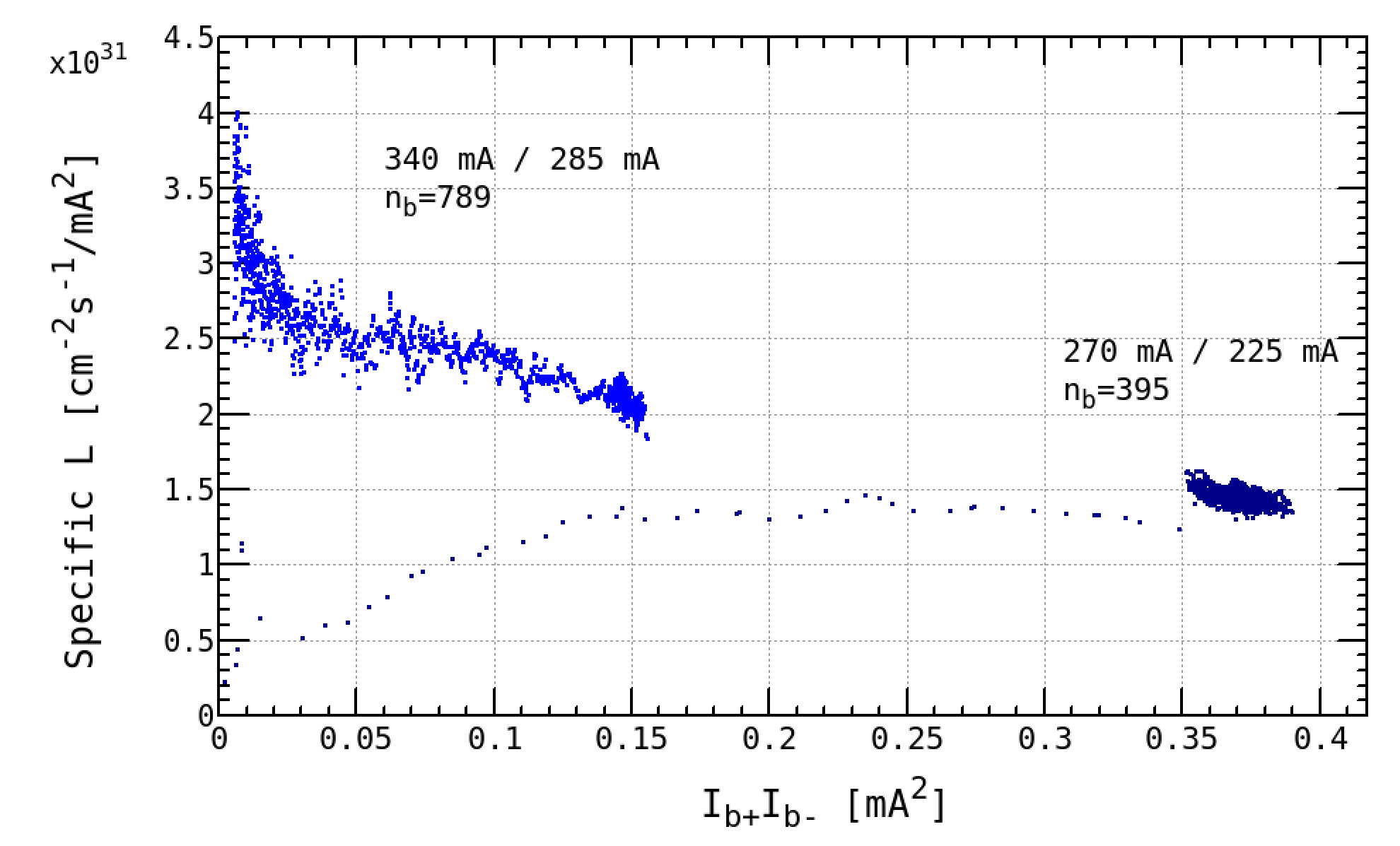}
   \caption{Specific luminosity as a function of bunch current product.}
   \label{fig:specLum}
\end{figure}

The machine parameters in Phase 2 and comparisons with the final parameters in Phase 3 are shown in Table~\ref{tab:machine}.  
The beam operations are classified according as ``High bunch current'', ``Reference'', and ``High current''.
In the high bunch current, the luminosity can be predicted to be 9$\times$10$^{33}$ \lum
if the number of bunches increases 4 times large with keeping the bunch currents.
The specific luminosity as a function of bunch current product for the reference and high bunch current is shown 
in Fig.~\ref{fig:specLum}.
If the lower bunch current product is focused, the specific luminosity is 4$\times$10$^{31}$ \slum~ 
which corresponds to $\bar{\sigma}_y^*$ = 300 nm ($\varepsilon_y$ = 30 pm).
The vertical beam size at the IP is consistent with the estimation from projected emittance, $\varepsilon_y$ = 23 pm, which 
measured by the X-ray beam size monitor in the LER.
However, the beam blowup was observed in the higher bunch current product.
The beam-beam simulations(weak-strong) suggest some indications that XY couplings, chromatic XY couplings, and 
skew sextupole components at the IP affect the beam-beam blowup.
This issue is under study and will be investigated in Phase 3.

\begin{figure}[htb]
   \centering
   \includegraphics[width=240pt]{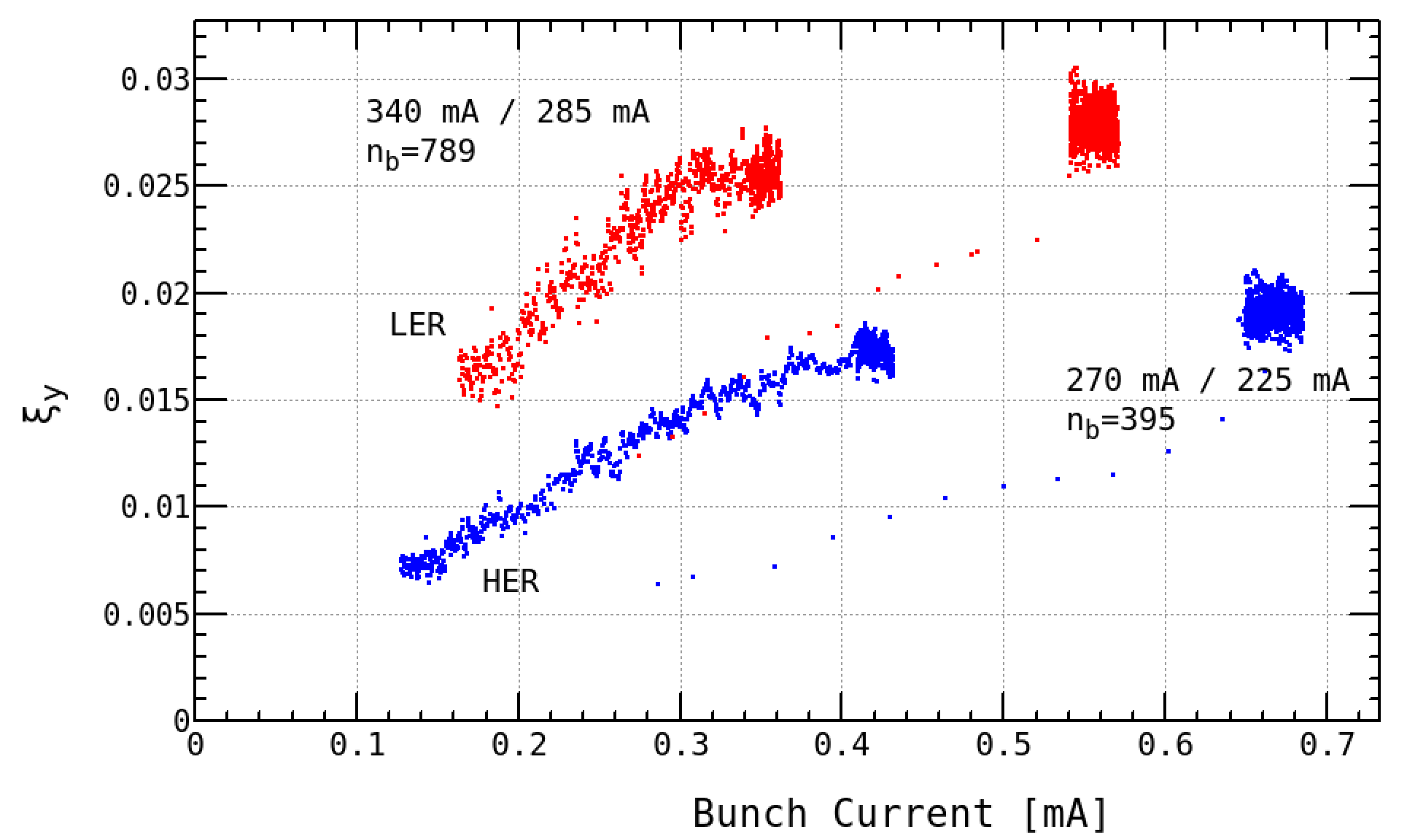}
   \caption{Beam-beam parameter as a function of bunch current product.}
   \label{fig:beambeam}
\end{figure}

\begin{figure}[htb]
   \centering
   \includegraphics[width=240pt]{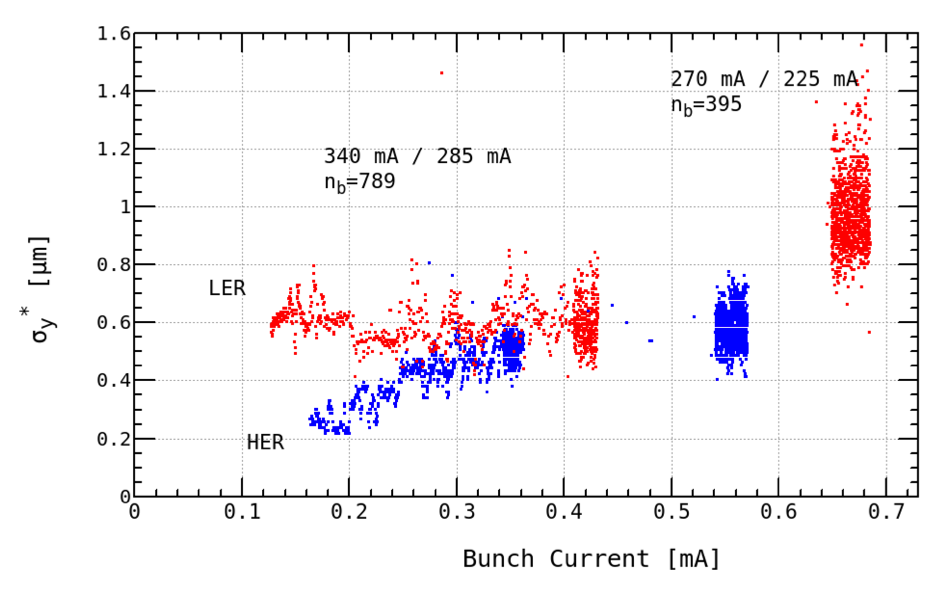}
   \caption{Vertical beam size at the IP as a function of bunch current. The beam size is measured by the X-ray beam size monitor.}
   \label{fig:sigma_y}
\end{figure}

Figure~\ref{fig:beambeam} shows the beam-beam parameters correspond to the specific luminosity as shown in Fig.~\ref{fig:specLum}.
Here, the beam-beam parameter is defined by
\begin{eqnarray}
\xi_{y\pm}=\frac{r_e N_\mp}{2\pi\gamma_\pm\sigma_z\phi_x}\frac{\beta_y^*}{\bar{\sigma}_y^*},
\end{eqnarray}
where $\bar{\sigma}_y^*$ is obtained from the measured specific luminosity as described in Eq.~\ref{eq:speclum}.
We observed the beam-beam parameter was saturated in the high bunch current.
This is a similar behavior of the specific luminosity obviously.
The vertical beam size at the IP is shown in Fig.~\ref{fig:sigma_y}.
The $\sigma_y^*$ is estimated by using X-ray  beam size monitor for each ring.
The beam blowup was clearly observed in the HER.
However, $\sigma_y^*$ in the LER was almost constant in the lower bunch current from 0.1 mA to 0.4 mA 
and was observed significant beam blowup in the high bunch current.
This implies that there is still machine error such as the XY couplings in the LER
which causes the sudden luminosity degradation in the small bunch current product less than 0.02 mA$^2$.
\begin{table*}[hbt]
   \centering
   \caption{Machine Parameters in Phase 2. 
     The parameters in Phase 3 is the final design of SuperKEKB.
     The $\sigma_y^*$ is estimated from the vertical beam size at the light-source point measured by the X-ray beam size monitor.
     The $\bar{\sigma}_y^*$ = $\Sigma_y^*/\sqrt{2}$ is obtained from the luminosity.
     Intra-beam scattering is considered in the horizontal emittance.
   }
   \begin{tabular}{l|cc|cc|cc|cc|c}
       \toprule
        &
       \multicolumn{6}{c|}{\textbf{Phase 2}}  & 
       \multicolumn{2}{c|}{\textbf{Phase 3}}  &  \\
        &
       \multicolumn{2}{c|}{High bunch current}  & 
       \multicolumn{2}{c|}{Reference}  &
       \multicolumn{2}{c|}{High current}  &
       \multicolumn{2}{c|}{Final}   & Unit\\
       
       &
       LER & HER & LER & HER & LER & HER & LER & HER & \\
       \midrule
       $I$ at $L_{peak}$ & 265 & 217 & 327 & 279 & 788 & 778 & 3600 & 2600 & mA \\
       $n_b$            & \multicolumn{2}{c|}{395} & \multicolumn{2}{c|}{789} &
                          \multicolumn{2}{c|}{1576} & \multicolumn{2}{c|}{2500} & \\
       $I/n_b$          & 0.670 & 0.549 & 0.414 & 0.353 & 0.500 & 0.494 & 1.44 & 1.04 & mA \\ 
       $\varepsilon_x$  & 1.8 & 4.6 & 1.7 & 4.6 & 1.7 & 4.6 & 3.2 & 4.6 & nm \\
       $\beta_x^*$      & 200 & 100 & 200 & 100 & 200 & 100 & 32 & 25 & mm \\                   
       $\beta_y^*$      & 3 & 3 & 3 & 3 & 3 & 3 & 0.27 & 0.3 & mm \\                   
       $\sigma_z$       & \multicolumn{8}{c|}{6} & mm \\
       $2\phi_x$        & \multicolumn{8}{c|}{83} & mrad \\
       $\Phi$           & 13.1 & 11.6 & 13.5 & 11.6 & 13.5 & 11.6 & 24.6 & 23.2 & \\ 
       $\nu_x$          & 44.562 & 45.542 & 44.558 & 45.541 & 44.561 & 45.545 & 44.53 & 45.53 & \\
       $\nu_y$          & 46.617 & 43.609 & 46.615 & 43.610 & 46.614 & 43.612 & 46.57 & 43.57 & \\
       $\sigma_y^*$     &  883 & 652 & 692 & 486 & 1285 & 528 & 48 & 62 & nm \\                    
       $\bar{\sigma}_y^*$ & \multicolumn{2}{c|}{797} & \multicolumn{2}{c|}{552} &
                               \multicolumn{2}{c|}{879} & \multicolumn{2}{c|}{55} & nm \\
       $\xi_y$          & 0.030 & 0.021 & 0.0277 & 0.0186 & 0.0244 & 0.0141 & 0.088 & 0.081 & \\                   
       $L_{sp}$         & \multicolumn{2}{c|}{1.57$\times$10$^{31}$} &
                          \multicolumn{2}{c|}{2.27$\times$10$^{31}$} & \multicolumn{2}{c|}{1.43$\times$10$^{31}$} &
                          \multicolumn{2}{c|}{2.14$\times$10$^{32}$} & cm$^{-2}$s$^{-1}/mA^2$ \\
       $L$              & \multicolumn{2}{c|}{2.29$\times$10$^{33}$} &
                          \multicolumn{2}{c|}{2.62$\times$10$^{33}$} & \multicolumn{2}{c|}{5.55$\times$10$^{33}$} &
                          \multicolumn{2}{c|}{8$\times$10$^{35}$} & cm$^{-2}$s$^{-1}$ \\

       \bottomrule
   \end{tabular}
   \label{tab:machine}
\end{table*}

\begin{figure}[htb]
   \centering
   \includegraphics[width=240pt]{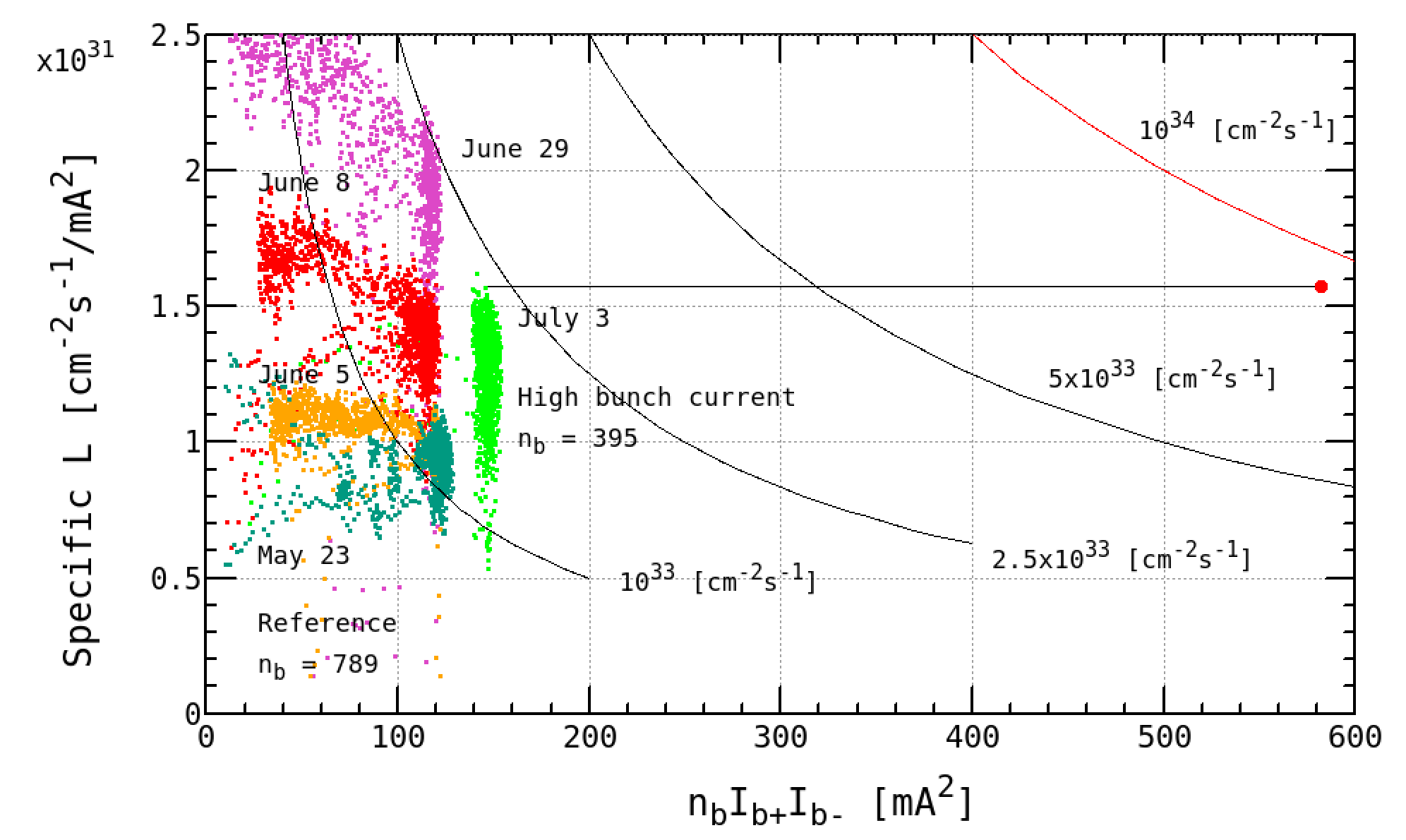}
   \caption{
     Specific luminosity as a function of bunch current product multiplied by number of bunches.
     Red point indicates the luminosity extrapolated from the high bunch current by multiplying factor of 4.
   }
   \label{fig:splum-lum}
\end{figure}

Figure~\ref{fig:splum-lum} shows the specific luminosity as a function of bunch current product multiplied by number of bunches.
It is found that the specific luminosity for the reference is improving day by day.
The total luminosity contours are also plotted in this figure.
The green plots corresponds to the high bunch current can be extrapolated to almost $L$ = 10$^{34}$ \lum by multiplying factor 4
as explained previously. 
The extrapolated beam current becomes 1060 mA with keeping the bunch current in the LER.

Since the beta squeezing was the first priority in Phase 2, it was focused for two months. 
The beam currents increased for about last one month.
The peak luminosity of 5.55$\times$10$^{33}$ \lum was achieved during the high current operation. 
However, the vertical emittance in the LER was increased by vertical dispersions artificially made according to dispersion knob
to increase Touscheck lifetime as much as possible.
Note that the peak luminosity is not optimized since the vertical beam size is large in the LER as shown in Table~\ref{tab:machine}.

\subsection{Electron Cloud}
\begin{figure}[htb]
   \centering
   \includegraphics[width=240pt]{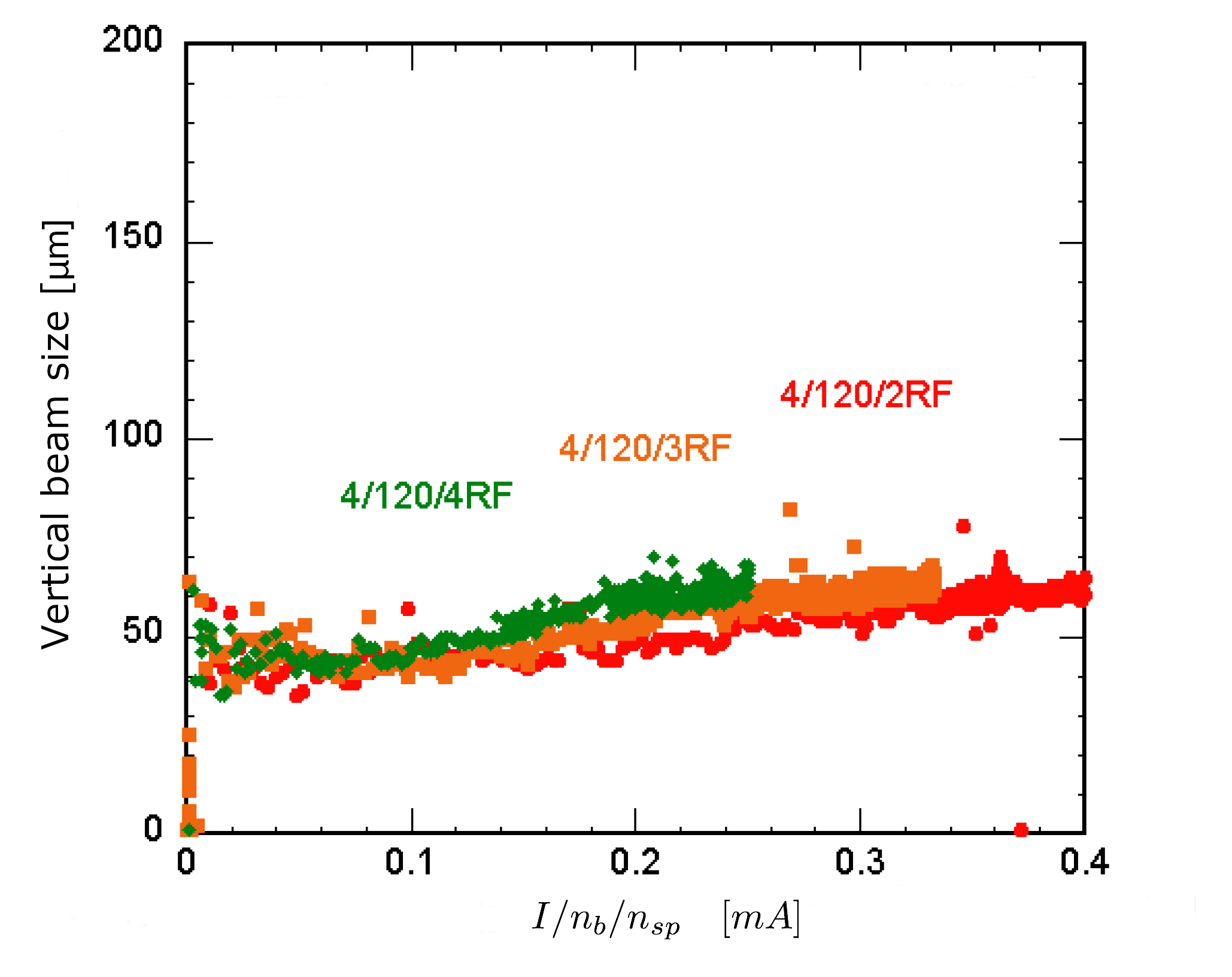}
   \caption{Vertical beam size measured by X-ray beam size monitor as a function of bunch current divided by rf bucket spacing.
   There are 3 fill patterns which are indicated by (number of trains)/$n_b$/$n_{sp}$.}
   \label{fig:ECI}
\end{figure}

The electron cloud effect(ECE)\cite{ref:ECE} was observed in Phase 1 
although the ante-chambers and TiN coating were adopted in the LER.
Therefore, additional solenoid-like permanent magnets have been installed for the beam pipes as much as possible 
before the Phase 2 commissioning since the electron cloud are produced and formed in the drift space.
Those magnetic field is several ten gausses.
Figure~\ref{fig:ECI} shows the vertical beam size as a function of bunch current divided by the rf bucket spacing.
The vertical beam size was measured by the X-ray beam size monitor in the LER.
The beam blowup due to ECE was not observed up to 0.4 mA as $I/n_b/n_{sp}$.
The threshold is much improved more than twice 0.2 mA which is the threshold observed in Phase 1.
The mode of coupled bunch instability changes and the growth rate is reduced after the installation of additional
permanent magnets. 

\section{Issues in Phase 2}
\subsection{QCS Quench}
We had 24 times QCS quenches during the Phase 2 commissioning in total.
Half of them occurred before the middle of April since the movable masks were almost fully open.
Beam loss of injection beams caused the most of QCS quenches before optics corrections after the beta squeezing. 
The horizontal oscillation of the injected beam due to injection error is transformed to the vertical oscillation 
because the XY coupling is very large before optics corrections.
Then, the beam hits the QCS because the vertical physical aperture in the QCS is smallest in the ring.
We decided to adjust the movable collimators not only to reduce the Belle II background but also
to avoid QCS quenches.
Several QCS quenches occurred until the end of May although the collimators were optimized, 
however, human error in the operation or troubles of the injection kickers caused them.
Another cure is the fast abort system with the diamond sensor has been adopted since the end of May.
No QCS quench occurred for a month since then.
Several QCS quenches occurred from the end of June to July again when the beam currents increased larger than 500 mA
and $\beta_y^*$ was 3 mm.
There were incidents that a head of the movable collimator was damaged with the QCS quench simultaneously.
The cases of QCS quench are categorized into during injection and during beam storage.
About 8000 particles hitting the superconducting coil causes a QCS quench when a simple calculation is considered
with an assumption of all energy lost.
However, more particles should be necessary because of an energy spread of the lost particles in the real machine.
The movable collimators and the fast abort system could avoid QCS quenches for the injection beams.
The both devices could avoid most of the QCS quenches for the storage beams, 
there were still a few events not understood.  
\subsection{Damage of Movable Collimator Head}
The vertical collimator\cite{ref:Vacuum} which is based on the design for PEP-II at SLAC has been installed for each ring.
We had a damage of the head in the vertical collimator for the LER and HER, respectively. 
We observed sudden pressure rise in the vicinity of the movable collimator accompanied by a beam abort and QCS quench. 
The injection background became very high and there was no way to reduce backgrounds by optimizing the collimator aperture
after this incident.
Thus, a gutter and spine on the collimator head were found, which were made by hitting the beams.
The reason of incident is unclear because beam instabilities and a large orbit drift were not observed.
There is a possibility of a dust trapping.
In order to cure the injection background due to the spine, the collimator was moved by a few mm in the horizontal direction.
Further investigation of the incident is necessary in Phase 3.  
\section{Conclusions}
We have performed the Phase 2 commissioning since 19th March until 17th July in 2018.
The beam currents are stored up to 860 mA in the LER and 800 mA in the HER, respectively.
We changed the machine parameters from the non-collision optics to the collision optics which is completely different lattice.
The beta functions at the IP were adopted to be 200 mm in the horizontal and 8 mm in the vertical plane 
for the initial collision tuning.
Then, we observed the first hadronic event on 26th April 2018.
The vertical beta function at the IP was successfully squeezed from 8 mm down to 3 mm.
The luminosity has increased even though $\beta_y^*$ is smaller than the bunch length.

We have experienced the issues such as QCS quenches and damages of collimator head.
We do not understand completely all of them and there are still unclear for some of the incidents.

We confirmed the nano-beam scheme in the Phase 2 commissioning and determined to move on Phase 3.  
\section{acknowledgment}
The authors wish to thank all of the Beast II groups and also the collaborators from overseas.
Many thanks go to the KEKB review committee for always encouraging us and suggesting plenty of recommendations for many years.  
%
%
%
%

%
\iffalse  
	\newpage
	\printbibliography

\else



\fi

\end{document}